\input harvmac

\input epsf
\ifx\epsfbox\UnDeFiNeD\message{(NO epsf.tex, FIGURES WILL BE
IGNORED)}
\def\figin#1{\vskip2in}
\else\message{(FIGURES WILL BE INCLUDED)}\def\figin#1{#1}\fi
\def\ifig#1#2#3{\xdef#1{fig.~\the\figno}
\goodbreak\topinsert\figin{\centerline{#3}}%
\smallskip\centerline{\vbox{\baselineskip12pt
\advance\hsize by -1truein\noindent{\bf Fig.~\the\figno:} #2}}
\bigskip\endinsert\global\advance\figno by1}

\def\ass{$A_4^{sl(2)}$}

\def \ha  { {\textstyle{1\ov 2} } }

\def \a {\alpha}

\def\bs{\bigskip }

\def\us{s_{pq}}
\def\ut{t_{pq}}
\def\uu{u_{pq}}
\def\asl{A_{4}^{sl(2)}(s,t)}

\def\slz{$SL(2,{\bf Z})$ }

\def \sm {\smallskip }

\def \ov {\over }
\def \four{{\textstyle{1\over 4}}}

\def \lr { \lref}
\def\np {{  Nucl. Phys. }}
\def \pl {{  Phys. Lett. }}

\def \ijmp {{ Int. J. Mod. Phys. }}

\baselineskip8pt
\Title{
\vbox
{\baselineskip 6pt{\hbox{ }}{\hbox
{UB-ECM-PF-04-36}}{\hbox{hep-th/0411035}} {\hbox{
  }}} }
{\vbox{\centerline {Effects of D-instantons in string amplitudes}
}}
\vskip -27 true pt
\centerline  {  Jorge G. Russo }

\bigskip

 \centerline {\it  Instituci\' o Catalana de Recerca i Estudis Avan\c{c}ats (ICREA),}
\smallskip
\centerline {\it  Departament ECM,
Facultat de F\'\i sica, Universitat de Barcelona,  Spain. }

\medskip\bigskip

\centerline {\bf Abstract}
\medskip
\baselineskip10pt
\noindent

We investigate the different energy regimes in the conjectured \slz invariant  four graviton scattering
amplitude
that incorporates D-instanton contributions in 10d type IIB 
superstring theory. We show 
that the infinite product over  \slz rotations  is convergent
in the whole complex plane $s,t$.
For high  energies $\a' s\gg 1$, fixed scattering angle,
and very weak coupling $g_s\ll 1/(\a' s)$, the four-graviton amplitude exhibits the usual exponential suppression.
As the energy approaches $1/g_s$, the suppression gradually diminishes until there appears a strong 
amplification near
a new pole coming from the exchange of a $(p,q)$ string. 
At energies $\a' s\ll 1/\sqrt{g_s} $,  
the pure D instanton contribution to the scattering amplitude is found
to produce a factor $A_4^{\rm D inst}\cong \exp (c
g_s^{3/2}e^{-{2\pi\over g_s}} s^3)$. At energies 
$1/\sqrt{g_s}\ll\a' s\ll 1/g_s $,  the
D-instanton factor becomes $A_4^{\rm D inst}\cong \exp (2
e^{-{2\pi\over g_s}+\pi g_s s^2})$,\ $\a'=4$.
At higher energies $\a's\gg 1/g_s$ the D-instanton contribution
becomes very important, and one finds an oscillatory behavior which alternates
suppression and amplification.
 This suggests that 
non-perturbative effects can lead to a high-energy behavior which is
significantly different from the perturbative string behavior.

\medskip
\Date {November 2004}
\noblackbox
\baselineskip 14pt plus 2pt minus 2pt

\lr \grgu{M.~B.~Green and M.~Gutperle,
``Effects of D-instantons,''
Nucl.\ Phys.\ B {\bf 498}, 195 (1997),
hep-th/9701093.
}

\lr \gv{M.B. Green and P. Vanhove, \pl B408 (1997) 122,
hep-th/9704145.}

\lr\mgreen{M.~B.~Green,
``Interconnections between type II superstrings, M theory and N = 4
Yang-Mills,''
hep-th/9903124.
 }

\lr \ggv{ M.~B.~Green, M.~Gutperle and P.~Vanhove,
``One loop in eleven dimensions,''
Phys.\ Lett.\ B {\bf 409}, 177 (1997).
}

\lr\rutse{
J.~G.~Russo and A.~A.~Tseytlin,
``One-loop four-graviton amplitude in eleven-dimensional supergravity,''
Nucl.\ Phys.\ B {\bf 508}, 245 (1997).}

\lr\jrusso{J.G. Russo, \pl B417 (1998) 253, hep-th/9707241.}

\lr\constru{J.~G.~Russo,
``Construction of SL(2,Z) invariant amplitudes in type IIB superstring
theory,''
Nucl.\ Phys.\ B {\bf 535}, 116 (1998),
hep-th/9802090.}

\lr \gvh{M.~B.~Green and P.~Vanhove,
``The low energy expansion of the one-loop type II superstring amplitude,''
Phys.\ Rev.\ D {\bf 61}, 104011 (2000),
hep-th/9910056.
}

\lr\kwon{M.~B.~Green, M.~Gutperle and H.~h.~Kwon,
Phys.\ Lett.\ B {\bf 421}, 149 (1998).}

\lr\GRGU{M.~B.~Green and M.~Gutperle,
``D-instanton partition functions,''
Phys.\ Rev.\ D {\bf 58}, 046007 (1998),
hep-th/9804123.
}

\lr\vanh{M.~B.~Green, H.~h.~Kwon and P.~Vanhove, 
``Two loops in eleven dimensions,''
 Phys.\ Rev.\ D {\bf 61}, 104010 (2000).
}

\lr\kiri{E.~Kiritsis and B.~Pioline,
Nucl.\ Phys.\ B {\bf 508}, 509 (1997),
hep-th/9707018.
}
\lr\berko{
N.~Berkovits,
Nucl.\ Phys.\ B {\bf 514}, 191 (1998),
hep-th/9709116.
}
\lr\kepa{  A.~Kehagias and H.~Partouche,
``D-instanton corrections as (p,q)-string effects and non-renormalization
theorems,''
Int.\ J.\ Mod.\ Phys.\ A {\bf 13}, 5075 (1998),
hep-th/9712164.
}

\lr\chalmer{G.~Chalmers and K.~Schalm,
JHEP {\bf 9910}, 016 (1999);
G.~Chalmers,
Nucl.\ Phys.\ B {\bf 580}, 193 (2000).
}

\lr\terras{A. Terras, {\it Harmonic Analysis on Symmetric Spaces and Applications}, vol. I, Springer-Verlag (1985).}

\lr\wittenb{E.~Witten,
``Bound states of strings and p-branes,''
Nucl.\ Phys.\ B {\bf 460}, 335 (1996).
}

\lr \john{J.H. Schwarz, \pl B360 (1995) 13,
hep-th/9508143.}

\lr\johns{Schwarz, old SL(2)..}

\lr\rusty{J.G. Russo and A.A. Tseytlin, \np B490 (1997) 121.}

\lr\iengo{R.~Iengo and C.~J.~Zhu,
JHEP {\bf 9906}, 011 (1999);
R.~Iengo,
JHEP {\bf 0202}, 035 (2002).
}

\lr \acv{ D. Amati, M. Ciafaloni and G. Veneziano, \ijmp  3 (1988) 615;
 \np B347 (1990) 550.}

\lr\Gross{D.~J.~Gross and P.~F.~Mende,
``The High-Energy Behavior Of String Scattering Amplitudes,''
Phys.\ Lett.\ B {\bf 197}, 129 (1987).
}

\lr\towns{P.~K.~Townsend,
``Membrane tension and manifest IIB S-duality,''
Phys.\ Lett.\ B {\bf 409}, 131 (1997)
hep-th/9705160;
M.~Cederwall and P.~K.~Townsend,
``The manifestly Sl(2,Z)-covariant superstring,''
JHEP {\bf 9709}, 003 (1997), hep-th/9709002.
}


\lr\grow{D.J. Gross and E. Witten, \np B277 (1986) 1.}

\lr\haro{S.~de Haro, A.~Sinkovics and K.~Skenderis,
``A supersymmetric completion of the R**4 term in IIB supergravity,''
Phys.\ Rev.\ D {\bf 67}, 084010 (2003),
hep-th/0210080.
}

\lr\mende{P.~F.~Mende and H.~Ooguri,
Nucl.\ Phys.\ B {\bf 339}, 641 (1990).}

\lr\bevaf{N.~Berkovits and C.~Vafa,
``Type IIB R**4 H**(4g-4) conjectures,''
Nucl.\ Phys.\ B {\bf 533}, 181 (1998),
hep-th/9803145.
}

\lr\venez{G.~Veneziano,
``String-theoretic unitary S-matrix at the threshold of black-hole
production,''
hep-th/0410166.
}



\def\brs{\bar s}
\def\brt{\bar t}
\def\bru{\bar u}

\def\bs{\bigskip }

\newsec{Introduction}

A  problem of interest is understanding what are the concrete effects 
that non-perturbative corrections can have in superstring theory, in particular, how they affect the high-energy behavior of string amplitudes.
In ten-dimensional type IIB superstring theory, the source of non-perturbative corrections
are the D instantons.

Computing the contribution of multiply-charged D instantons directly is complicated.
However, combining different pieces of information,  Green and Gutperle \grgu\  conjectured 
 the exact modular function that multiplies the $R^4$ term in the type IIB effective action,
 which exactly incorporates the infinite set of multiply-charged D instanton corrections.

One of the constraints on the effective action used by Green and Gutperle 
is precisely \slz\ invariance.
The \slz\ symmetry of  type IIB superstring theory 
 requires that the effective action must be invariant under $SL(2,{\bf Z})$ transformations 
to all orders in the $\a' $ expansion.
In particular, this implies that  graviton scattering amplitudes must be $SL(2,{\bf Z})$ invariant,
since there is a direct correspondence between the terms in the effective action and
the momentum expansion of the scattering amplitude.

In \refs{\jrusso,\constru} an \slz invariant four graviton amplitude was constructed by
applying a simple \slz\ symmetrization of the tree-level string theory
four graviton amplitude. The construction follows essentially the same rule used
by Green and Gutperle to symmetrize
the $R^4$ term.
It was conjectured that this scattering amplitude incorporates the
full series of D instanton corrections with the different D-instanton numbers.
This symmetric amplitude satisfies a number of  consistency conditions. In particular,
 corrections of perturbative origin appear
with an integer power of $g_s^2$. This is non-trivial and 
does not hold for any symmetrization.
It is also consistent  with the conjecture that 
high derivative terms in the type II effective action of the form $H^{4k-4}R^4$
should not receive  perturbative contributions beyond genus $k$ \bevaf . 
By construction, it reproduces the exact $R^4$ term proposed in \grgu ,
and it can be viewed as  a tree-level amplitude that accounts for the exchange of 
$(p,q)$ string states \wittenb .

These $(p,q)$ string states have a simple eleven-dimensional origin \john .
Type IIB superstring theory is obtained from M theory by compactification on a
2-torus and taking the zero area limit at fixed torus moduli.
In this limit, most membrane states get an infinite mass, except a certain set
of states that represent the $(p,q)$ strings of uncompactified 10d type IIB string theory.
These states are precisely the states that contribute as simple poles
in the \slz invariant amplitude of \refs{\jrusso,\constru}.

In this work we investigate the properties of the \slz invariant amplitude.
In particular, we factorize the pure D instanton contribution and study the high energy limit.

\phantom{\gv\ggv\rutse\kiri\kwon\berko\kepa\bevaf\GRGU\mgreen
\iengo\gvh\vanh\haro}

The conjecture of \grgu\  has withstood different tests and has been 
generalized in different directions  \refs{\gv -\haro }.
The idea of organizing  type IIB perturbation theory in \slz\ invariant way
was also suggested by \refs{\towns , \chalmer}.
Scattering amplitudes at high energies incorporating higher genus
effects were investigated by \Gross\  and \acv .

\newsec{\slz invariant amplitude}

The four-graviton scattering amplitude for 10d type IIB superstring introduced 
in \refs{\jrusso,\constru}\ is given by the following formula:
\eqn\vvzz{
A_4=\kappa^2  K \asl \ ,
}
\eqn\vvx{
\asl ={1\ov s t  u} \prod_{(p,q)'}
{\Gamma (1- \us)\Gamma (1-\ut)
\Gamma(1-\uu )\ov \Gamma (1+ \us )\Gamma (1+\ut )
\Gamma(1+\uu ) }\ ,
}
\eqn\stuu{
\us ={\a' s\ov 4|p+q\tau |}\ ,\ \ \ \ut ={\a' t\ov 4|p+q\tau |}\ ,\ \ \ 
\uu ={\a' u\ov 4|p+q\tau |}\ ,\ \ \ \ \us+\ut+\uu =0\ ,
}
where  $p$ and $q$ are relatively prime,
$\tau=C^{(0)}+i g_s^{-1}$ is the usual coupling of type IIB superstring theory, and
$K$ is the same kinematical factor depending on the momenta and 
polarization of the external
states appearing in the tree-level 
Virasoro amplitude  (see e.g. \grow )
$$
K=\zeta^{AA'}_1\zeta^{BB'}_2\zeta^{CC'}_3\zeta^{DD'}_4  K_{ABCD}(k_i)K_{A'B'C'D'}(k_i)\ ,\ 
$$
$$
 K_{ABCD}=-{1\ov 4}s t \ \eta_{AC}\eta_{BD}+...
$$
The scattering amplitude \vvx\ can also be written as
\eqn\vrz{
 \asl ={1\ov stu } e^{\delta (s,t)} \ ,
}
with
\eqn\ddelta{
\delta (s,t) = 2 \sum _{k=1}^\infty {\zeta (2k+1) g_s^{k+1/2}
E_{k+1/2}(\tau )\ov 2k+1} 
\big(\brs ^{2k+1}+\brt ^{2k+1}+\bru ^{2k+1} \big)\ ,
}
$$
\bar s=\four \a' s\ ,\  \ \ \bar t=\four \a' t\ , \ \ \ \bar u=\four
\a' u\ ,\ \ \ \ \brs+\brt+\bru =0\ ,
$$
and $E_r(\tau )$ is the non-holomorphic Eisenstein series, given by (${\rm Re}\  r>1$)
\eqn\ess{
E_r(\tau )=\sum_{(p,q)'} {\tau_2^r\ov |p+q\tau |^{2r} }\ .
}
{}For very small coupling $g_s\ll 1$, the terms with $q\neq 0$ are negligible in 
the sum \ddelta , so that $g_s^{k+1/2}E_{k+1/2}(\tau )\to 1$. One recovers the  tree-level
four-graviton Virasoro amplitude,
\eqn\vrzo{
A_4(s,t)=\kappa^2  K A_4^0(s,t) \ ,\ \ \ \ \ A_4^0 (s,t)={1\ov stu } e^{\delta _0(s,t)} \ ,
}
\eqn\logve{
\delta _0(s,t)= 2\sum_{k=1}^{\infty }
{\zeta (2k+1)\ov 2k+1} \big(\brs ^{2k+1}+\brt ^{2k+1}+\bru ^{2k+1} \big)\ .
}
This gives
\eqn\eex{
A_4^0(s,t) ={1\ov s t  u} 
{\Gamma (1- \bar s)\Gamma (1-\bar t)
\Gamma(1-\bar u )\ov \Gamma (1+ \bar s )\Gamma (1+\bar t )
\Gamma(1+\bar u ) }\ .
}
The scattering amplitude \ass \ adds to the Virasoro amplitude perturbative and non-perturbative  contributions.
They are seen explicitly by expanding the Eisenstein functions 
at large $\tau _2 =g_s^{-1}$ \terras ,
\eqn\bbb{
E_r(\tau )=\tau_2^r+\gamma_r \tau_2^{1-r}+
{4\tau_2^{1/2}\pi^r\ov\zeta(2r)\Gamma(r)}
\sum_{n,w=1}^\infty \big({w\ov n}\big)^{r-1/2}\cos(2\pi  wn\tau_1)K_{r-1/2}(2\pi w n\tau_2 )\ ,
}
$$
\gamma_r={\sqrt{\pi }\ \Gamma(r-1/2)\ 
\zeta(2r-1)\ov \Gamma(r)\ \zeta(2 r) }\ .
$$
Using the asymptotic expansion for the Bessel function $K_{r-1/2}$,
\eqn\asbe{
K_{r-1/2}(2\pi w n\tau_2 )={1\ov \sqrt{4wn\tau_2 } }e^{-2\pi w n\tau_2}\sum_{m=0}^\infty {1\ov (4\pi wn \tau_2)^m }{\Gamma(r+m)\ov \Gamma(r-m)m! }\ ,
}
we see that the $E_{k+1/2}(\tau )$ terms in the amplitude
are of the form
\eqn\qqq{
g_B^{k+1/2} E_{k+1/2}(\tau )=1+\gamma _{k+1/2}\ g^{2k}_B
+O\big( e^{-2\pi/g_B}\big)\ .
}
Note that the
 non-perturbative contributions are $O(e^{-{2\pi m\over g_s}})$, where $m=wn$ is an integer number. 
The coefficient $2\pi m $  is  crucial
in order to have a one-to-one correspondence between these terms
and instanton contributions. It is a remarkable fact that the product over  \slz\ rotations 
automatically generates the full series of D-instanton contributions.

\smallskip

We summarize the main  properties of \ass :
\smallskip

\noindent 1)  It is $SL(2,{\bf Z})$ invariant. This is explicit in the Einstein frame,
$g_{\mu\nu}^E=g_B^{-1/2} g_{\mu\nu}$, so that $s_E=g_B^{1/2} s$, $t_E=g_B^{1/2} t$, $u_E=g_B^{1/2} u$,
and $s_E, t_E, u_E$ remain fixed under \slz\ transformations.

\smallskip

\noindent 2) It adds perturbative $g_s^{2k}$ and non-perturbative $O(e^{-{2\pi m \over g_s}})$
corrections to the  Virasoro amplitude.
\sm

\noindent 3) It has simple poles in the $s$-$t$-$u$ channels at $\us=n $, $\ut=n $,
$\uu=n $, $n=0,1,2,...$
corresponding to a tree-level exchange of particles with masses
\eqn\polos{
\four \a' M^2 =n |p+q\tau |\ .
}

\smallskip

\noindent 4) It reproduces the exact  (proportional to $E_{3/2}(\tau )$) 
$R^4 $ term conjectured in \refs{\grgu  }, containing a one-loop correction and the  full D-instanton contributions.
It also reproduces the exact $\zeta(5)E_{5/2}(\tau )\nabla^4 R^4$ term conjectured in \vanh \ (moreover, in \gvh\ there
was a  calculation of a genus one term in $\nabla^6 R^4$ which was found proportional to $2\zeta(3)\zeta(2)$, 
which differs from the prediction of the \ass\ amplitude only by a factor of 2).


\medskip

The spectrum \polos\  is  the  spectrum of  $(p,q)$ string states \wittenb :
\eqn\ppqq{
 M^2=4\pi T_{pq}(N_R+N_L)={2\ov\a'} \ |p+q\tau |\ (N_R+N_L)\ ,\ \ \ \ N_R=N_L\ .
}
This spectrum corresponds to the zero winding sector of the spectrum
studied in  \refs{\john ,\rusty} for the nine-dimensional type IIB
string theory. Setting $\tau_1=C^{(0)}=0$, the full spectrum in $D=9$ is given by
\eqn\ppdd{
 M_9^2={n^2\over R_{10}^2} \big(p^2+{q^2\over g_s^2}\big)+{w_{10}R_{10}\over {\a'}^2}
+ {2\ov\a'} \ \sqrt{p^2+{q^2\over g_s^2}}\ (N_R+N_L)\ ,\ \ \ \ N_R-N_L=n w_{10}\ ,
}
where
$R_{10}$ is the radius of the compact tenth dimension. In the limit $R_{10}\to \infty $,
one must set, as usual, the winding number $w_{10}$ to zero to have finite mass. The term proportional to
${1\over R_{10}^2}$ becomes the continuous 10d component of the momentum $p_{10}$, so that $M^2=M^2_9-p_{10}^2$
and one gets eq. \ppqq .
It is important to note that a charged D string has mass $M=O(1/g_s)$, as seen from \ppdd .
The neutral $(p,q)$ strings of ten dimensions have masses given by \ppqq\ of order $M=O(1/\sqrt{g_s})$ 
for $q\neq 0$. This is why the product over \slz rotations produces poles at $\a' s=O(1/g_s)$. 

The collection of states \ppqq\  are the only quantum states of M-theory compactified
on a 2-torus
that remain of finite mass after taking the zero-area limit of the torus that leads to
ten-dimensional type IIB string theory \constru . 
The scattering amplitude \ass\ can thus be viewed as a tree-level scattering amplitude where all these states are
exchanged.

\smallskip

The scattering amplitude \ass\ does not describe loop effects such as 
discontinuity cuts (see \refs{\constru,\gvh } for  discussions).
In particular, it should  not be a good approximation of the full 
scattering amplitude at $g_s=O(1)$ and $\a´s$ large.
It represents an improvement of the Virasoro amplitude at $g_s\ll 1$ 
(or at the S-dual situation, $g_s\gg 1$),
where D-instanton (and some perturbative) contributions have been incorporated.

\smallskip

The exchange of $(p,q)$ string states is clear from the pole structure of \vvx .  It becomes
manifest by writing 
 \eqn\esta{
\delta  = \ha \sum_{(m,n)\neq (0,0)}  \log {M^2_{mn}+ s\ov M^2_{mn}- s}
\  +(s\to t)+ (s\to u)\ ,
}
where
$$
\a' M_{mn}^2=4 |m+n\tau |\ .
$$
This generalizes the analogous formula for the Virasoro amplitude, with $\delta_0 $ written in the form
\eqn\suit{
\delta_0 = \sum_{m=1}^\infty \delta _{(m)} \ ,
\ \ \ \delta _{(m)}= \log {M_m^2+ s\ov M_m^2- s}
\  +(s\to t)+ (s\to u)
\ , \ \ \ \a' M_m^2=4m \ .
}
Now the sum in \esta\ contains not only the terms $\a' M_m^2=4 m$, but all the terms $ M_{mn}^2$,
representing all $(p,q)$ string states.

\newsec{ Convergence properties }

The scattering amplitude \ass\ is defined through an infinite product \vvx\ over a pair $(p,q)$
of relatively  prime integers, i.e. integers $(p,q)$ having greatest common divisor equal to one.
An important issue is what are the convergence properties of this product.

To study the convergence, we write
\eqn\cutf{
\delta (s,t) =\sum_{(p,q)'} \log { \Gamma(1-s_{pq}) \over \Gamma(1+s_{pq}) } +(s\to t)+(s\to u) \ .
}
{}We have to look at the behavior of terms with large $p,q$.
For any given $s,t, u$, there are positive  integers $(p_0,q_0)$ such that
all  terms with  $p>p_0,\ q>q_0$ have $|p+q\tau |\gg s,t,u$ and $s_{pq},\ t_{pq},\ u_{pq}$
small. These terms have the behavior
\eqn\sutf{
\log { \Gamma(1-s_{pq}) \Gamma(1-t_{pq})
\Gamma(1-u_{pq})\over \Gamma(1+s_{pq})\Gamma(1+t_{pq}) \Gamma(1+u_{pq})}
\cong {2\zeta(3)\over 3|p+q\tau |^{3}}\big(\bar s^3+\bar t^3+\bar u^3) \ ,
}
where we have used $s+t+u=0$. 
The sum over $p,q$ of $|p+q\tau |^{-3}$ is known to be convergent \terras\ 
(in particular, the full sum over $p,q$ of $|p+q\tau |^{-3}$ 
defines $E_{3\over 2}(\tau )$, see \ess ).
Therefore the sum  in \cutf\  is convergent.
Since we have made no assumption about $s,t$, the series \cutf\  has infinite radius of convergence.

We have also investigated the convergence numerically, by 
explicit
calculation of the infinite product
in different sectors of
the complex planes $s$ and $t$, for generic values of the coupling $\tau $.
As an additional check, we have also computed the amplitude in the representation \esta ,
obtaining the same (finite) numerical results.


Note that the series  \ddelta\  defining the amplitude has a finite radius of convergence
if we write $\zeta(2k+1)=\sum_m m^{-2k-1}$ and perform first the sum over $k$.
The sum over $k$ is the series of a logarithm and it diverges when $s,\ t$ or $u$ meet the first
pole.
The same of course applies for the Virasoro amplitude written in the form \logve .

\newsec{ Approach to the different scales }

Let us assume $g_s\ll 1$, and examine the different scales that appear as 
the center-of-mass energy is increased
from zero. We set $C^{(0)}=\tau_1=0$, so that $\tau=i\tau_2=ig_s^{-1}$.

\medskip

\subsec{Region $\a' s\ll {1\over g_s}$}

 For $\a' s\ll 1$, one has $\asl \to {1\over stu}$, and one recovers the supergravity tree-level four graviton amplitude.

\medskip

 In general,  for any $\a' s\ll {1\over g_s}$, 
one has $\asl \to A_4^0(s,t)$, one recovers the Virasoro amplitude \eex\ with its usual properties: 
 simple poles at $\bar s =n$, $\bar t =n$, $\bar  u =n$, with $n$ positive 
 (recall that in the physical region of elastic scattering $s$ is positive, while $t,u$
are negative).
The high energy behavior is as follows:
\smallskip

\noindent i) High $\a' s$, fixed scattering angle $\varphi $: 
\eqn\fxx{
\asl \to A_4^0(s,t)\cong {1\over stu} \ e^{-\a ' a_0\ s }\ ,\ \ \ 
}
$$
a_0={1\over 2} \big| \sin^2{\varphi\over 2}\log \sin^ 2{\varphi\over 2}+\cos^ 2{\varphi\over 2}\log \cos^ 2{\varphi\over 2}\big| \  ,
$$
$$
t=-s \sin^2{\varphi\over 2}\ ,\ \ u=-s  \cos^ 2{\varphi\over 2} \ .
$$

\smallskip

\noindent ii) High $\a' s$, fixed $t$: 
 \eqn\cnz{
\asl \to A_4^0(s,t) \cong {1\over stu} (-1)^t\ s^{-{\a'\over 2}|t|} \ 
{\Gamma(1-{\a' t\over 4})\over \Gamma(1+{\a't\over 4})}\ .
}
\medskip

It is useful to split the scattering amplitude \ass\ in three factors:
\eqn\aamh{
\asl =A_4^0\times A_4^{\rm pert} \times A_4^{\rm D inst}={1\over stu} \ e^{\delta_0+
\delta_{\rm pert}+
\delta_{\rm Dinst}}\ .
}
Here $A_4^{\rm pert} $ represents perturbative corrections to the Virasoro amplitude
of the form $g_s^{2k}$ coming from the term $\gamma_r\tau_2^{1-r}$ in the expansion \bbb .
The remaining factor $A_4^{\rm D inst}$ represents the pure D-instanton contribution, 
 terms proportional to $e^{-2\pi wn/g_s}$
coming  $K_{r-1/2}$ in \bbb . They are given by
\eqn\pertu{
\delta_{\rm pert}=\sqrt{\pi }\sum_{k=1}^\infty {(k-1)!\zeta(2k)\over \Gamma(k+{3\over 2})}g_s^{2k}
( \bar s^{2k+1}+ \bar t^{2k+1}+ \bar u^{2k+1})\ ,
}
\eqn\instanton{
\delta_{\rm Dinst}= {4\sqrt{\pi}}\sum_{n,w,k=1}^\infty  \big({w\ov n}\big)^{k}
{\pi^kg_s^k\ov \Gamma(k+{3\over 2})}
K_{k}\big( {2\pi w n\over g_s} \big) ( \bar s^{2k+1}+ \bar t^{2k+1}+ \bar u^{2k+1}) \ .
}
These series converge for $\bar s< {1\over g_s}$. 
In the region $\bar s\ll {1\over g_s}$, the leading behavior of $\delta_{\rm pert}$
is just given by the first term in the 
series \pertu , $\delta_{\rm pert}\cong {2\pi^2\over 9}( \bar s^{3}+ \bar t^{3}+ \bar u^{3})$.

Assuming $g_s\ll 1$, one can use the asymptotic form of the Bessel function \asbe .
Then one obtains that in this region  $\a' s\ll {1\over g_s}$, the D-instanton contribution is  given by
\eqn\instant{
\eqalign{
\delta_{\rm Dinst} &= 2\bar  s \sqrt{\pi g_s} e^{-{2\pi\over g_s}}
\sum_{k=1}^\infty  {(\pi g_s\bar  s^2)^k\ov \Gamma(k+{3\over 2})}\ +(s\to t)+(s\to u)
\cr
&=2 {\rm sign }(s) e^{-{2\pi\over g_s}}e^{\pi g_s \bar s^2}{\rm Erf}(\sqrt{\pi g_s \bar s^2})\ +(s\to t)+(s\to u)\ ,
}}
where Erf is the error function.
There are two regimes, $\bar s\ll 1/\sqrt{g_s}$, so that  $\pi g_s \bar s^2\ll 1$, and
$ 1/\sqrt{g_s}\ll \bar s\ll 1/g_s$.
In the first case, we get 
\eqn\instaa{
\delta_{\rm Dinst}\cong {8\pi\over 3} g^{3/2}\ e^{-{2\pi\over g_s}}  \big( \bar s^{3}+ \bar t^{3}+ \bar u^{3}\big)\ , \ \ \ 
\ \bar s\ll 1/\sqrt{g_s}\ . 
}
This is a positive contribution in the physical region of the Mandelstam parameters, but it is
negligible compared to $\delta_0 $ and 
$\delta_{\rm pert}$.
In the second case, we get
\eqn\instaa{
\delta_{\rm Dinst}\cong 2e^{-{2\pi\over g_s}}  \big( e^{\pi g_s \bar s^2}-e^{\pi g_s \bar t^2}-e^{\pi g_s \bar u^2}\big)\ ,
\ \ \ \ 1/\sqrt{g_s}\ll \bar s\ll 1/g_s\ .
}
This is still  tiny, since in this region ${\pi g_s \bar s^2}\ll 2\pi/g_s$~.

\subsec{   Region $\a' s=O(1/g_s)$}

 In this case one begins to see simple poles at 
$\a' s=\sqrt{m^2+n^2/g_s^2}$ with $n\neq 0$. 
Since $g_s$ is small, there
is an accumulation of poles with $m=0,1,2,...$ near the pole at $n=1$, at $n=2$, etc. 
These poles are not seen in a coarse grain plot of the amplitude, since
they appear at  special points.

\ifig\fone{$\delta (s,t)$ as a function of $\bar s={\a' s\over 4}$ at fixed scattering
angle $\varphi ={\pi\over 2}$ for $g_s=0.01$. The straight line is $\delta_0 (s,t)$.
}
{\epsfxsize=8.0cm \epsfysize=5.0cm \epsfbox{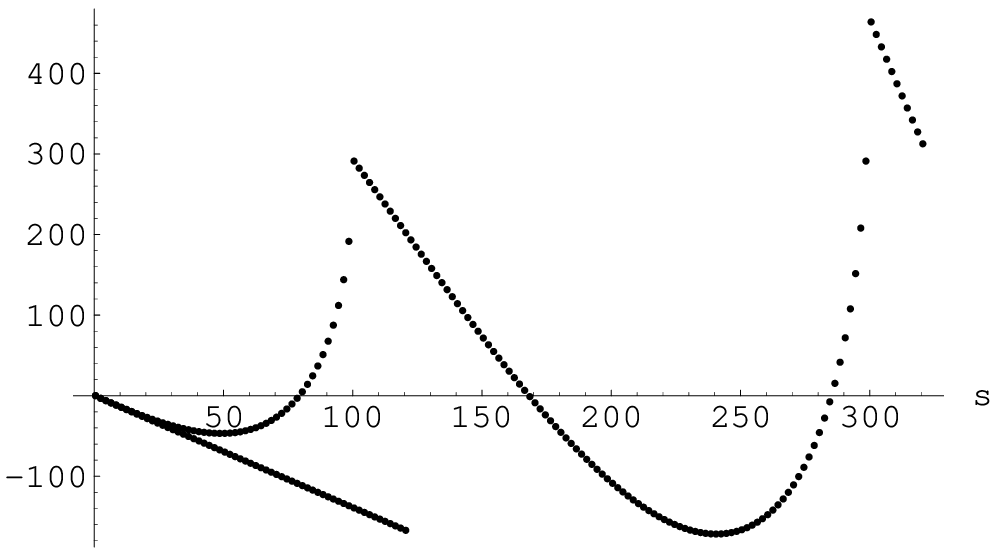}}

Figure 1 shows $\delta (s,t)$ 
(the logarithm of the amplitude, see \vrz ) as a function of $s$ for large $s,t,u$ and 
fixed scattering angle $\varphi ={\pi\over 2}$ and for $g_s=0.01$.
One can see that at the beginning there is the straight line with negative slope as in \fxx , reproducing
the usual 
suppression of the Virasoro amplitude. 
Then $\delta (s,t)$ becomes positive, producing an amplification of the amplitude near $\bar s=1/g_s$ 
(see figure 2).
As $s$ is further increased, the amplitude diminishes;  then it is amplified again near $\bar s=3/g_s$.

The behavior can be understood as a combination of the effects of $\delta_0$ and $\delta_{\rm pert}$, since
the D-instanton contribution $\delta_{\rm Dinst}$ is still negligible in this region for $g_s=0.01$.
Figure 2 shows the two contributions separately.

\ifig\fdos{The  separate contributions to \ass\ for $g_s=0.01$:
a) The tree-level Virasoro part $\delta_0 $ is a straight line.
b) The perturbative part $\delta_{\rm pert}$ has cusps and is positive, 
giving an amplification effect. c) The D instanton part $\delta_{\rm Dinst}$ 
is still negligible at $\bar s<320$.
}
{\epsfxsize=7.0cm \epsfysize=5.0cm \epsfbox{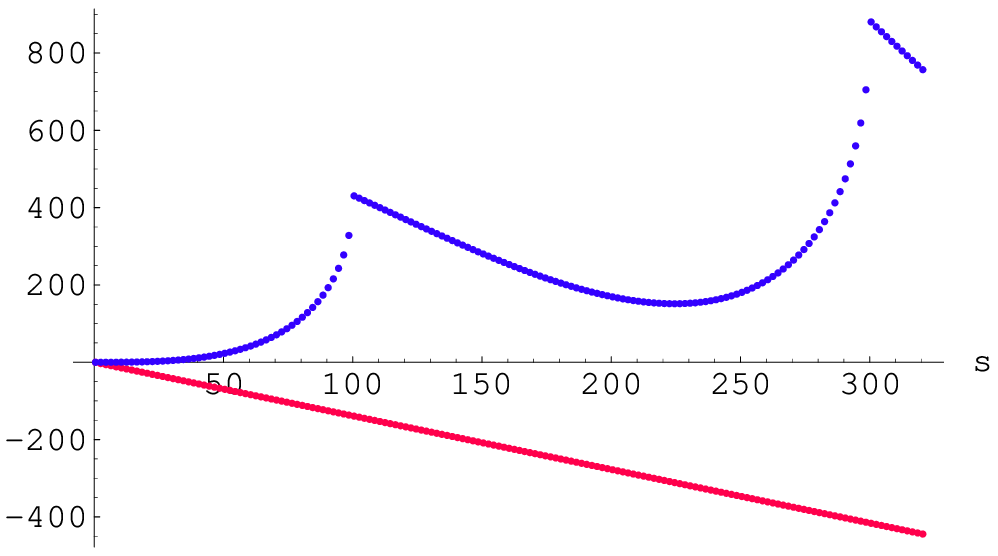}}

\subsec{Region $\a' s\gg {1\over g_s} $}

To examine the behavior in this region, we again consider the three contributions
$\delta_0,\ 
\delta_{\rm pert} ,\ 
\delta_{\rm Dinst}$ separately.

The perturbative part \pertu\ can be resummed explicitly, with the result \constru :
\eqn\pret{
\delta_{\rm pert}=-4\sum_{m=1}^\infty \sqrt{{m^2\over g_s^2}-\bar s^2} \arcsin{\bar s g_s\over m}
+(s\to t)+(s\to u)\ .
}
Its high energy behavior is shown in  Figure 3, which indicates a behavior $\delta_{\rm pert}\cong {\rm const.}\ s$.
More precisely, it is bounded between two straight lines: $1.22\bar s<\delta_{\rm pert}<1.69\bar s$.
On the other hand, we have from \fxx :
\eqn\virlaw{
\delta_{0}\cong -\a' a_0\ s\ .
}
\ifig\fcua{The perturbative part  $\delta_{\rm pert}$ computed at ultra high energies.
It grows linearly with $s$.
}
{\epsfxsize=8.0cm \epsfysize=5.0cm \epsfbox{ 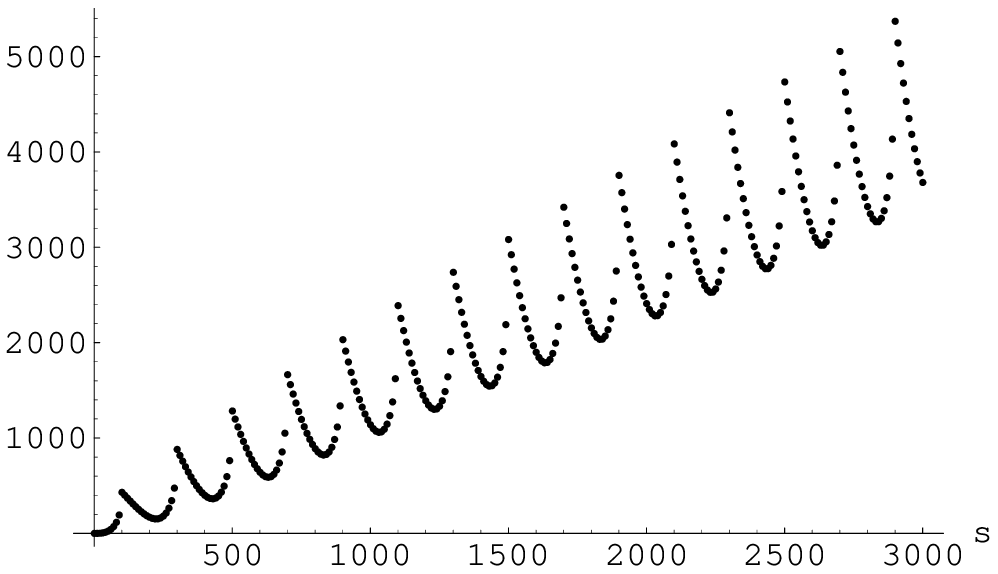}}

The pure D instanton part $\delta_{\rm Dinst}$ can be computed 
from $\delta_{\rm Dinst}=\delta - \delta_0-
\delta_{\rm pert}$, where $\delta (s,t)$ and $\delta_{\rm pert}$ are
 computed from the convergent
sums \cutf\ and \pret .
Numerically, one finds that $\delta_{\rm Dinst}$
 oscillates between negative and positive values, which are
of the same order of magnitude as  $\delta_0, \delta_{\rm pert}$.
This gives rise to a behavior which alternates strong 
suppression and amplification of the amplitude as $s$ is increased.

The asymptotic behavior at very large $s$ is unclear
since numerical precision is worst at high $s$.
It is plausible that at $\a' s\gg 1/g_s$  there are higher genus corrections
 not contained in \ass\ which become important.
 Among the different types of corrections, there are gravitational 
corrections corresponding to multiple 
exchange of gravitons \refs{\acv,  \venez }.
 In the present case of high energy and
fixed scattering angle, the dominant genus $h$ contribution is known
\Gross , though it is unclear how to resum the full series \mende .

We find remarkable that the product over \slz rotations produces a
convergent, 
mathematically well 
defined amplitude,
and that the  infinite D-instanton sum produces significant changes
in the high energy behavior. It would be  interesting
to understand how to incorporate higher genus corrections to \ass\ in an
\slz invariant way.


\bs\bs

\noindent{\bf Acknowledgements}
\medskip

We would like to thank  P. Vanhove  for  useful remarks.
We also thank CERN for hospitality where this work was carried out
This work is partially supported by MCYT FPA,
2001-3598 and CIRIT GC 2001SGR-00065, and
by the European Community's Human potential
Programme under the contract HPRN-CT-2000-00131.


\listrefs

\vfill\eject\end